\documentclass[preprint, showpacs]{revtex4}

\usepackage{epsfig}
\usepackage{amsmath}

\newcommand{\be}{\begin{equation}}
\newcommand{\ee}{\end{equation}}

\newcommand{\RW}{\mathrm{RW}}
\newcommand{\AM}{\mathrm{AM}}
\newcommand{\kB}{k_\mathrm{B}}

\newcommand{\bd}[1]{\mathbf{#1}}

\newcommand{\vrp}{\bd{r}'}

\newcommand{\dd}{\mathrm{d}}
\newcommand{\rme}{\mathrm{e}}
\newcommand{\rmi}{\mathrm{i}}

\begin{document}

\title{Comment on "Casimir force acting on magnetostatic bodies embedded in media"}

\author{Iver Brevik\footnote{E-mail: iver.h.brevik@ntnu.no} and Simen A. Ellingsen\footnote{E-mail: simen.a.ellingsen@ntnu.no}}
\affiliation{Department of Energy and Process Engineering, Norwegian University
of Science and Technology, N-7491 Trondheim, Norway}

\begin{abstract}
In a recent paper [C. Raabe  and D.-G. Welsch, Phys. Rev. A
\textbf{71}, 013814 (2005)] an electromagnetic energy-momentum
tensor is suggested as an alternative to the Abraham-Minkowski
tensor and is applied to calculations of Casimir forces
 in planar geometries. We argue that the universality of the suggested  tensor is
 doubtful; application  of the Raabe-Welsch theory to a simple  example in  classical
 electrodynamics shows that their proposed tensor is unable to
 describe the situation in a simple way.
 We also show that modified Casimir forces acting on the cavity medium as prescribed by these authors suffer from
   problems of definiteness and peculiar properties which call for experimental support before this theory can be regarded as  acceptable.%deleted "eventually"
\end{abstract}

\pacs{42.50.Wk, 42.50.Nn, 12.20.-m}

\maketitle

\section{Introduction}

In a recent paper \cite{raabe05}, Raabe and Welsch suggested a theory for
electromagnetic fields in a medium based on the Lorentz
force, from which they derived a stress tensor different from the
Abraham-Minkowski tensor traditionally applied in systems where
electro- and magnetostrictive forces may be neglected.
%Added:
The same tensor was
first proposed by Poincelot many years ago \cite{poincelot67}, and has more
recently been supported by Obukhov and Hehl \cite{obukhov03}.

The Casimir effect, electromagnetic forces originating from the
fluctuating electromagnetic fields in vacuum, was chosen by
these authors as the physical phenomenon to which to apply the new
theory. This is a surprising choice because the long standing debate over the
correct energy-momentum tensor has essentially been one of
classical electrodynamics where a number of experiments have been
performed in order to distinguish between candidates
\cite{brevik79}. Experimental progress in the field of Casimir
forces on the other hand is relatively recent, and no experimental measurement of
the Casimir force to date can distinguish between the %two stress tensors.
different stress tensors.
Because compared to a
number of effects in optics and classical electrodynamics the
Casimir effect is imperfectly understood, formally complicated and
experimentally difficult to access, we argue that the study of Casimir
phenomena is, at least at present, ill-suited as arena for the debate over
which electromagnetic stress tensor, or more generally, energy-momentum tensor, is the more appropriate.

In the only criticism of the Raabe and Welsch theory published so
far \cite{pitaevskii06}, Pitaevskii countered Raabe and Welsch,
%demonstrating, convincingly we find, that
%CHANGE:
arguing that
the stress tensor
proposed by these authors is but a part of the complete stress
tensor (cf. also the Reply in Ref.~\cite{raabe06r}). Where
Pitaevskii's arguments were of %a phenomenological
a theoretical
nature, we will
start the other end and analyse the consequences of using the
newly proposed tensor.

The stress tensor in a medium (traditionally defined as the negative of the space components of the four dimensional energy-momentum tensor) hereafter called the Raabe-Welsch tensor - is
\be\label{TRW}
  T_{ik}^\RW = \epsilon_0E_iE_k + \frac{1}{\mu_0}B_iB_k - \frac{1}{2}\delta_{ik}(\epsilon_0\bd{E}^2 +
  \frac{1}{\mu_0}\bd{B}^2).
\ee
 We will
use SI units throughout. In an isotropic medium the Abraham and
Minkowski stress tensors become equal and we may write it as
\be\label{TAM}
  T_{ik}^\AM = E_iD_k + H_iB_k - \frac{1}{2}\delta_{ik}(\bd{E}\cdot\bd{D} +
  \bd{H}\cdot\bd{B}).
\ee
In a linear magnetodielectric we have $\bd{D}=\epsilon_0\epsilon\bd{E}$ and $\bd{B}=\mu_0\mu\bd{H}$, where we define $\epsilon$ and $\mu$
 dimensionless, relative to their vacuum values. In vacuum the two stress tensors are obviously identical.

 We give also the corresponding expressions for the volume force
 densities, assuming static conditions. In standard notation,
 \begin{align}
 {\bf f}^\RW &= (\rho-{\bf \nabla \cdot P)}{\bf E}+{\bf J\times
 B+(\nabla \times M)\times B}, \label{3}\\
 {\bf f}^\AM &=\rho {\bf E}+{\bf J \times
 B}-\frac{\epsilon_0}{2}E^2{\bf
 \nabla}\epsilon-\frac{\mu_0}{2}H^2{\bf \nabla}\mu. \label{4}
 \end{align}
 Here $\rho$ and $\bf{J}$ are respectively the charge and current density due to external charges not accounted for by the polarisation and magnetisation fields.

 As mentioned, a survey was given by one of the present authors some years ago \cite{brevik79} on
 the electromagnetic energy-momentum in matter, emphasizing the
 possibilities for experimental discrimination between the
 different alternatives (cf. also Ref.~\cite{brevik86}). Since
 then, a number of papers have appeared on the energy-momentum
 problem and related matters; some of them are listed in
 Refs.~\cite{obukhov03,kentwell87,loudon97,antoci97,antoci98,garrison04,feigel04,loudon04,welsch05,leonhardt06,raabe06,birkeland07,nieminen07}.
 The recent review of Pfeifer {\it et al.} \cite{pfeifer07} also
 contains many references. Readers interested in an introduction to
 the electromagnetic energy-momentum problem, may consult the nice
 exposition in M{\o}ller's book \cite{moller72}.

\section{The Raabe-Welsch tensor applied to a simple electrostatic example}

The electromagnetic energy-momentum tensor describes a nonclosed
medium - the field itself plus its interaction with matter - and
some arbitrariness is therefore involved in choosing the proper
expression for it. An important condition on this tensor is that
it shall be able to describe the outcome of experiments in
classical electromagnetism in a simple and straightforward way.
%SENTENCE ADDED
According to a recent (re)analysis by Obukhov and Hehl \cite{obukhov03} the usefulness of the tensor (\ref{TRW}) is supported by the experiments of Walker and Walker \cite{walker75} and James \cite{james68}. In the following we discuss another experiment which indicates the opposite.

\begin{figure}[htb]
  \begin{center}
    \includegraphics[width=1in]{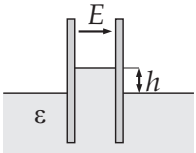}
    \caption{Condenser plates immersed in dielectric liquid.}\label{fig_condensator}
  \end{center}
\end{figure}

Consider the elementary electrostatic situation sketched in
figure \ref{fig_condensator}, namely two parallel metallic condenser plates partially
immersed in a dielectric liquid of  permittivity $\epsilon$. When
a horizontal electric field $\bf E$ is applied across the plates,
the liquid rises to some equilibrium height $h$. According to the
Abraham-Minkowski picture, the force density ${\bf f}^\AM
=-(\epsilon_0/2)E^2{\bf \nabla}\epsilon$ in the interior of the
homogeneous liquid will be {\it zero}. The elevation of the free
surface is  the result of the vertically directed component
$f_{z}^\AM$ acting in the boundary layer of the surface.
Integrating $f_{z}^\AM$ across the boundary layer, or
alternatively calculating the difference between the stress
components $T_{zz}^\AM$ on the two sides, one finds the surface
force density to be $(\epsilon_0/2)(\epsilon-1)E^2$. Equating this
to the gravity pressure $\rho gh$, one obtains for the height\cite{brevik79}
\begin{equation}
h=\frac{\epsilon_0}{2\rho g}(\epsilon-1)E^2. \label{5}
\end{equation}
From a physical point of view this is not the complete picture,
however; there act in addition {\it electrostrictive} forces in
the liquid which seek to compress the liquid in the interior
domain. Actually, this force has to be stronger than the force
acting in the free surface region; otherwise the liquid would not
be able to rise as a coherent body%. This aspect is discussed more closely in Ref.~\cite{brevik79}, page 144,
(\cite{brevik79} p.144). Electrostriction will not be further
considered here since it is not essential in the following; the
height in  Eq.~(\ref{5}) is found, as we have seen, as a
consequence of the net Abraham-Minkowski force only.

The question is now: How can the result of Eq.~(\ref{5}) be
explained by using the Raabe-Welsch tensor? According to
Eq.~(\ref{3}) it follows that, when $\rho={\bf J}=0$, ${\bf
f}^\RW$ is  zero in the homogeneous interior of the liquid where
${\bf \nabla \cdot P}=0$. In the boundary layer at the free
surface  $\bf \nabla \cdot P$ is not zero, but the force $ -(\bf
\nabla \cdot P)E$ can never be vertically directed all the time
that $\bf E$ is horizontal. Looking at it in another way, by
taking the difference between the stress components $T_{zz}^\RW$
on the two sides of the free surface, we find the same zero result
for the vertical force. We are therefore led to the conclusion
that for practical purposes, the Raabe-Welsch tensor is not very
appropriate to explain this simple experiment.

Let us emphasize: We are not here entering into a consideration of
the theoretical foundations of the Raabe-Welsch tensor. Our
purpose has merely been to investigate the practical usefulness of
it. %NOTE ADDED
(Note that the authors of \cite{raabe05}, after deriving the tensor (\ref{TRW}) very generally, applies it exclusively to the Casimir force \cite{raabe06r}, a quantum mechanical phenomenon in contrast with the classical situation considered above and, for example, in \cite{obukhov03}).

\section{The Casimir force as predicted by Raabe-Welsch}

\begin{figure}[htb]
  \begin{center}
    \includegraphics[width=1.7in]{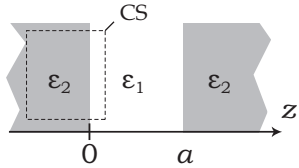}
    \caption{The standard three layered geometry with a control surface (CS) enclosing the leftmost interface.}\label{fig_3_layer}
  \end{center}
\end{figure}

To calculate Casimir forces by means of a given stress tensor, the quantum mean of the products of the electric and magnetic field squares are expressed in terms of the imaginary part of a Green's function through the fluctuation-dissipation theorem. The
Green's function of the system may in turn be readily determined in the $k_\perp, \omega, z$ Fourier domain in a
multilayered geometry by a classical electrodynamical theory of multiple scattering \cite{tomas95}, and upon insertion into the chosen stress tensor the Casimir force is calculated as a surface integral over a control surface enclosing the relevant interfaces of the system, as examplified in figure \ref{fig_3_layer}. In plane parallel systems with surfaces normal to the $z$ axis the force per unit transverse area is simply the difference between the $zz$ component of $\mathbf{T}$ immediately to the the right and left of the interface,
\be\label{pressureT}
  F = A_\perp(\langle T_{zz,\text{right}}\rangle - \langle T_{zz,\text{left}}\rangle)
\ee
where $\langle \cdots\rangle$ denotes the quantum mean.

In \cite{raabe05} the Casimir force on a plate inside a cavity is calculated in this way and further analysed by Toma\v{s} \cite{tomas05}. It may appear  surprising that this rather complicated 5-layered geometry is chosen, as it adds considerable mathematical complications as compared to the simpler three-layer geometry considered by Lifshitz \cite{lifshitz56} and later by numerous others. A slightly simpler geometry of a slab of finite width outside a half-space was subsequently considered \cite{raabe05b,tomas06} and it was shown that a modified Casimir-Polder force \cite{casimir48b} was obtained in the case where the slab is a slice of the same dilute medium as fills the interspace. While the five-layer geometry has also been considered in the Abraham-Minkowski formalism \cite{tomas02,ellingsen07,ellingsen07b} and has interesting properties, we will consider the Raabe-Welsch result for the simplest and most well-known 3-layer system shown in Fig. \ref{fig_3_layer}. This is for transparency and because no experiments have been performed in a sandwich geometry to date.

We assume the gap width to be $a$, $\mu=1$ everywhere, and the wall material to be equal on both sides for simplicity. Let furthermore the gap material be more dilute than the walls for imaginary frequencies, so that $\epsilon_2 > \epsilon_1~\forall \rmi\zeta$. We choose $z=0$ at the left interface.

In the three layered system the force is found by use of (\ref{pressureT}) on a control surface enclosing one of the interfaces; we choose the leftmost interface. The pressure in the gap, defined negative for attraction, is then the negative of the force on the left interface divided by transverse area,
\[
  P = \langle T_{zz,\text{left}}\rangle-\langle T_{zz,\text{gap}}\rangle.
\]

The Green's function plays the r\^ole of a propagator; the relative extent to which a source at $(\vrp, t')$ gives
   rise to a field at $(\mathbf{r}, t)$, $t>t'$. Mathematically, the Green's function in the  ($k_\perp, \omega; z,z'$) domain has terms dependent on $z-z'$ and $z+z'$, but upon insertion into the Abraham-Minkowski stress tensor  (\ref{TAM}) any term which is function of $z+z'$ vanishes, regardless of the properties of these terms (shown generally in \cite{ellingsenMasters}).

For this reason the Abraham-Minkowski tensor has two important
properties which we will see that the Raabe-Welsch tensor does not
share:
\begin{enumerate}
  \item $\langle T_{zz}\rangle$ does not depend on the position $z$ inside a homogeneous region.
  \item $\langle T_{zz} \rangle$ is zero in the regions outside the gap in the geometry of figure \ref{fig_3_layer}. This is because all terms of the Green's function are proportional to $z+z'$ for $z<0$ (and correspondingly for $z>a$) which is physically obvious as the Green's function would otherwise diverge away from the interfaces \cite{tomas95}.
\end{enumerate}
The Casimir force in the Abraham-Minkowski formalism is thus found simply as (the integral over all real frequencies becomes a sum of the imaginary Matsubara frequencies) $P^\AM = -\langle T_{zz,\text{gap}}^\AM \rangle$, i.e.\
\be
  P^\AM = -\frac{\kB T}{\pi} {\sum_{m=0}^\infty}' \int_0^\infty \dd k_\perp~k_\perp \kappa_1
  \sum_{q=p,s}\frac{1}{d_q}.\label{AMtensor}
\ee
The prime on the summation mark means that the $m=0$ term is
taken with half weight, s and p denote the TE and TM polarisation
respectively, and we use
\[
  \kappa_i = \sqrt{k_\perp^2 + \frac{\epsilon_i \zeta^2}{c^2}};~~\frac{1}{d_q} = \frac{r_q^2 \rme^{-2\kappa_1 a}}{1-r_q^2 \rme^{-2\kappa_1
  a}}, \quad i=1,2,
\]
where $r_q$ is the Fresnel reflection coefficient of the gap-wall
interfaces (the same for both interfaces),
\[
  r_s = \frac{\kappa_2 - \kappa_1}{\kappa_2 + \kappa_1};~~ r_p = \frac{\epsilon_1\kappa_2 - \epsilon_2\kappa_1}{\epsilon_1\kappa_2 + \epsilon_2\kappa_1}.
\]
We notice that $r_s\geq 0$ and $r_p\leq 0$. The expressions for
$d_q, r_s$  and $r_p$ are functions of the imaginary Matsubara frequency
$\omega=i\zeta_m=2\pi i\kB T m /\hbar$.

When inserted into the Raabe-Welsch tensor (\ref{TRW}), however, the $z+z'$-dependent terms of the Green's function do not cancel, meaning that $T_{zz}^\RW$ is $z$ dependent as $z\to z'$
%NOTE ADDED
(the authors of \cite{raabe05} presume without comment that the optical theory of \cite{tomas95} is consistent with their proposed theory).
Because of this the Casimir attraction between the two slabs separated by medium 1 is not well defined, since it depends on the exact position of the boundaries of the control surface over which the tensor is integrated.

Since $\langle T_{zz}^\RW\rangle$ outside the gap region decreases exponentially away from the interfaces, however, we assume that a workable procedure is to assume that the stress tensor outside the gap is evaluated at infinity and is therefore also zero, in which case the force per unit area acting on the interface inside the chosen control volume also has the form $|P^\RW(z)| = \langle T_{zz,\text{gap}}^\RW(z)\rangle$.

The expression obtained with the tensor due to Raabe and Welsch is far more complicated than (\ref{AMtensor}):
\begin{widetext}
  \be\label{RWforce}
    \langle T_{zz}^\RW(z) \rangle = \frac{\kB T}{\pi} {\sum_{m=0}^\infty}' \int_0^\infty\!\! \dd k_\perp~k_\perp \left\{\kappa_1 \!\!\sum_{q=p,s}\!\!(\delta_{sq}+\frac{1}{\epsilon_1}\delta_{pq})\frac{1}{d_q}
    - \frac{\epsilon_1\zeta_m^2}{2\kappa_1 c^2}\left(1-\frac{1}{\epsilon_1}\right)\sum_{q=p,s}(\delta_{sq}-\delta_{pq})\left[\frac{1}{d_q}+\chi_q(z)\right]\right\},
  \ee
\end{widetext}
where the dependence on $z$ is through the function
\[
  \chi_q(z) = \frac{r_q \rme^{-\kappa_1 a}}{1-r_q^2\rme^{-2\kappa_1 a}}\cosh 2\kappa_1(z-a/2).
\]

The fact that $\langle T_{zz}\rangle$ is $z$ dependent, say Raabe and Welsch, solves a ``paradox'' in the standard theory, that the force on a slice of material in the interspace vanishes when the Abraham-Minkowski approach is used. Interestingly, this is the exact opposite of what Pitaevskii and Lifshitz argue \cite{dzyaloshinskii61}: the fact that the terms dependent on $z+z'$ should vanish is, they assert, ``obvious from physical considerations [since] we should [otherwise] obtain a momentum flux in the gap that varied with the coordinate, which would contradict the law of conservation of momentum.'' Since Raabe and Welsch predict just such a spatially varying momentum flux, there must be balancing forces present which their stress tensor does not account for. Such balancing forces are not considered in \cite{raabe05}. The reply is that these ``additional (internal and external) forces'' are not taken into account in the force expression because ``only electromagnetic contributions are included in the stress tensor''\cite{raabe06r}. However, such contact forces ensuring mechanical equilibrium inside the medium are also of electromagnetic origin, and should thus be accounted for if \emph{all} forces stemming from Lorentz type interactions were truly included. To separate one from the other appears artificial and may not be helpful since it is normally experimentally unfeasible to measure one of a pair of mutually cancelling forces.

Another peculiar trait is that the force acting inside the homogeneous cavity medium itself as predicted by Raabe and Welsch is divergent when $z$ approaches the interfaces at $0$ and $a$. In this case the integrand of the $k_\perp$ integral in (\ref{RWforce}) is a nonzero constant as $k_\perp\to\infty$ and the integral diverges as a linear function of the the upper integral limit.

The authors recognise this and argue \cite{raabe05} that
the $k_\perp$ integral should in be cut off because
``large values of [$k_\perp$] correspond to very oblique
travelling waves ... [which] are not reflected but walk off
instead'' due to the finite transverse dimensions of any real
system. This argument is not correct in our opinion, although it
is true that the Casimir force may be regarded
as a sum of multiple reflections between the interfaces %, a fact which is drawn upon explicitly in deriving the Green's function \cite{tomas95} and is easy to recognize in the form of the Abraham-Minkowski expression (\ref{AMtensor})
\cite{ellingsen07b}.
To see how the argument is flawed we must turn to real frequencies for a moment since imaginary frequencies $\zeta$ have no obvious physical meaning. Assume for the sake of argument that $\epsilon_1$ is real and
positive for all (real) $\omega$. Then the longitudinal wave vector $k_z$ is real for
$k_\perp \leq \sqrt{\epsilon_1} \omega/c$ and imaginary for larger
values of $k_\perp$. The obliquely travelling waves of which Raabe
and Welsch speak are thus found in a small range of
$k_\perp$ values just below $\sqrt{\epsilon_1}\omega/c$, whereas
larger values represent evanescent fields which stay on the surfaces \footnote{In real systems,
$\epsilon$ has an imaginary part causing attenuation of
propagating fields in media}. The Abraham-Minkowski force integral
(\ref{AMtensor}) converges therefore, not because these
large-$k_\perp$ fields disappear beyond the system's periphery, but
because they are exponentially attenuated away from the surfaces too quickly to
 cause interaction across the gap. There is no reason that we can see to assume that a
cutoff of the integral to be appropriate.

%PARAGRAPH ADDED
One may note that the divergencies which appear in (\ref{RWforce}) when $z$ approaches $0$ or $a$ are formally
similar to those one obtains with the Lifshitz formula (\ref{AMtensor}) when letting the gap
 separation $a$ tend to zero. If, as proposed in \cite{raabe05b}, the Raabe Welsch force may be thought
  of as Casimir-Polder forces acting on particles in the interspace medium, such similarity could make
  physical sense since also the Casimir-Polder force diverges with vanishing separation. The divergence of
   (\ref{RWforce}) as $a\to 0$ can be seen as a breakdown of simplifying assumptions such as sharp boundaries
    in this limit whereas the physical source of the divergence of (\ref{RWforce}) as $z\to 0$ we find less obvious.
     Eq. (\ref{RWforce}) should intuitively represent the net sum of all Casimir forces acting on all particles enclosed
      by the control surface, yet counterintuitively the force \emph{increases} the less of the cavity medium is enclosed, and diverges in the limit where the surface encloses only the left hand medium.

This said, there remains at least one setting in which the new tensor could be of interest, namely systems such as figure \ref{fig_3_layer} in which the cavity medium is compressible and out of mechanical equilibrium, a setting the Abraham-Minkowski tensor does not cover since its derivation assumes mechanical equilibrium\cite{pitaevskii06}. A candidate could be a cavity filled with a polarizable gas, whose equilibrium configuration will be one in which the Casimir-Polder forces on the molecules increases the density of particles near the walls. Such a set-up could conceivably be employed to test the predictions of Raabe and Welsch.

\section{Conclusions}

We have shown that the stress tensor due to Raabe and Welsch does not stand up to the criterion that it should be able to explain simple electromechanical experiments in a straightforward manner, a reasonable criterion to apply to any candidate electromagnetic stress tensor. It is our opinion that the field of the Casimir effect, being a formally complicated, imperfectly understood and experimentally difficult area of physics, is ill suited for the introduction of a theory which pertains in essence to classical electrodynamics.

As applied to the Casimir force, the suggested tensor gives results which are mathematically difficult and which do not immediately yield a numerical prediction for forces which may be measured in a laboratory setting. While the theory is interesting and could be of value to Casimir experiments on systems out of mechanical equilibrium, the task remains for the proponents of this new tensor to suggest experiments which may demonstrate the novel insights attributed to it.


\begin{thebibliography}{99}
\bibitem{raabe05}
C. Raabe  and D.-G. Welsch, Phys. Rev. A {\bf 71}, 013814 (2005).
 \bibitem{poincelot67}
 P. Poincelot, C. R. Acad. Sci. Paris {\bf 264B}, 1064, 1179,
 1225, 1560 (1967).
 \bibitem{obukhov03}
 Y. N. Obukhov and F. W. Hehl, Phys. Lett. A {\bf 311}, 277 (2003)
\bibitem{brevik79}
I.  Brevik, Phys. Rep.  {\bf 52}, 133 (1979).
\bibitem{pitaevskii06}
L. P. Pitaevskii,  Phys. Rev. A {\bf 73}, 047801 (2006).
\bibitem{raabe06r}
 C. Raabe and D.-G. Welsch, Phys. Rev. A {\bf 73},
 047802 (2006).
\bibitem{brevik86}
I. Brevik, Phys. Rev. B {\bf 33}, 1058 (1986).
\bibitem{kentwell87}
G. W. Kentwell and D. A. Jones, Phys. Rep. {\bf 145}, 319 (1987).
\bibitem{loudon97}
R. Loudon, L. Allen, and D. F. Nelson, Phys. Rev. E {\bf 55}, 1071
(1997).
\bibitem{antoci97}
S. Antoci and L. Mihich, Nuovo Cim. B {\bf 112}, 991 (1997).
\bibitem{antoci98}
S. Antoci and L. Mihich, Eur. Phys. J. D {\bf 3}, 205 (1998).
\bibitem{garrison04}
J. C. Garrison and R. Y. Chiao, Phys. Rev. A {\bf 70}, 053826
(2004).
\bibitem{feigel04}
A. Feigel, Phys. Rev. Lett. {\bf 92}, 020404 (2004).
\bibitem{loudon04}
R. Loudon, Fortschr. Physik {\bf 52}, 1134 (2004).
\bibitem{welsch05}
D.-G. Welsch and C. Raabe, e-print arXiv quant-ph/0501170.
\bibitem{leonhardt06}
U. Leonhardt, Phys. Rev. A {\bf 73}, 032108 (2006).
\bibitem{raabe06}
C. Raabe and D.-G. Welsch,  e-print arXiv:quant-ph/0602059.
\bibitem{birkeland07}
O. J. Birkeland and I. Brevik, Phys. Rev. E {\bf 76}, 066605
(2007).
\bibitem{nieminen07}
T. A. Nieminen, V. L. Y. Loke, A. B. Stilgoe, G. Kn{\"o}ner, A. M.
Bra{\'n}czyk, N. R. Heckenberg, and H. Rubinsztein-Dunlop, J. Opt.
A: Pure Appl. Opt. {\bf 9}, S196 (2007).
\bibitem{pfeifer07}
R. N. C. Pfeifer, T. A. Nieminen, N. R. Heckenberg, and H.
Rubinsztein-Dunlop, Rev. Mod. Phys. {\bf 79}, 1197 (2007).
\bibitem{moller72}
C. M{\o}ller, {\it The Theory of Relativity}, 2nd ed. (Clarendon
Press, Oxford, 1972), Ch. 7.
\bibitem{walker75} G. B. Walker and D. G. Lahoz, Nature {\bf 253}, 339 (1975); G. B. Walker and G. Walker, Nature {\bf 263}, 401 (1976); {\bf 265}, 324 (1977).
\bibitem{james68} R. P. James, Proc. Natl. Acad. Sci. USA {\bf 1}, 1149 (1968).


%section 3
\bibitem{tomas05}
 M. S. Toma\v{s},  Phys. Rev. A {\bf 71}, 060101(R) (2005); Fizika A {\bf 14} 29 (2005)
\bibitem{lifshitz56} E. M. Lifshitz, Sov. Phys. JETP {\bf 2} 73 (1956)
\bibitem{raabe05b} C. Raabe and D.-G. Welsch, J. Opt. B:  Quantum Semiclass. Opt. {\bf 7} S610 (2005)
\bibitem{tomas06} M. S. Toma\v{s}, J. Phys. A {\bf 39}, 6785 (2006)
\bibitem{casimir48b} H. B. G. Casimir and D. Polder, Phys. Rev. {\bf 73} 360 (1948)
\bibitem{tomas02} M. S. Toma\v{s}, Phys. Rev. A {\bf 66}, 052103 (2002)
\bibitem{ellingsen07}
 S. A. Ellingsen, J. Phys. A {\bf 40}, 1951 (2007)
\bibitem{ellingsen07b}
 S. A. Ellingsen and I. Brevik, J. Phys. A {\bf 40}, 3643 (2007)
\bibitem{tomas95}
 M. S. Toma\v{s}, Phys. Rev. A {\bf 51}, 2545 (1995).
\bibitem{ellingsenMasters}
  S. A. Ellingsen, Master's thesis, Department of Physics, Norwegian University of Science and Technology, Trondheim, Norway, 2006 (unpublished) Appendix B.4
\bibitem{dzyaloshinskii61}
I. E.  Dzyaloshinskii, E. M. Lifshitz, and L. P. Pitaevskii, Sov.
Phys. Usp. {\bf 4}, 153 (1961); E. M.  Lifshitz  and L. P. Pitaevskii, {\it  Statistical Physics Part 2} (Pergamon Press, Oxford, 1981), \S 81.

\end{thebibliography}
\end{document}